\documentclass[superscriptaddress,prl,floatfix,twocolumn,amsmath,amssymb,aps,nofootinbib,showkeys,showpacs]{revtex4-1}

\usepackage{graphicx}
\usepackage{amsmath}
\usepackage{amssymb}
\usepackage{dcolumn}
\usepackage{color}
\usepackage{multirow}
\usepackage{bm}
\usepackage{subfigure}
\usepackage{tabularx}
\usepackage{booktabs}
\usepackage{siunitx}
\usepackage{mhchem}

\begin{document}
\title{Switchable polarization manipulation, optical logical gates and conveyor belt based on U-shaped $\ce{VO2}$ nanoholes}
\author{Xiao-Yu Ouyang}
\email{ouyangxiaoyu@stu.pku.edu.cn}
\affiliation{School of Physics, Peking University, Beijing 100871, P. R. China}
\affiliation{Yuanpei College, Peking University, Beijing 100871, P. R. China}
\author{Jing Yang}
\email{jingyang@pku.edu.cn}
\affiliation{School of Physics, Peking University, Beijing 100871, P. R. China}
\date{\today}

\begin{abstract}
    Based on U-shaped plasmonic nanoholes in an $\ce{Au-VO2-Au}$ film, we propose to achieve several switchable functions at the telecom wavelength by transition from the $\ce{VO2}$ semiconductive state to the metallic state. The first is the polarization manipulation of four different polarization states ($x\&y$-polarization, LCP, and RCP). An array of U-shaped holes constitutes of the high efficiency SPP splitter, and thus the spin-encoded optical logical gates can be achieved. Furthurmore, we prove that a nano-optical conveyor belt can be build up with such U-shaped holes, making the transport of nanoparticles over the film efficiency by transforming between two spin states periodically, and the transport direction switchably along with the $\ce{VO2}$ phase.
\end{abstract}


\maketitle

\section{Introduction}
Surface Plasmon Polatiron (SPP) has been widely studied and applied in the past decades. Manipulating the polarization state of light using different shape of nanoholes is one of the applications, especially multifunctional manipulating with materials.  It is worth mentioning that $\ce{VO2}$ has distinctive phase change characteristics because it shows a semiconducter-to-metal transition around room temperature (~68°), which gives convenience for practical applications in switchable functions for controling the SPPs.
\par
In this work, the switchable splitter for either $x\&y$ polarization (linear state) or LCP/RCP (spin state) is proposed using a U-shaped nanohole in hybrid $\ce{Au-VO2-Au}$ films. The multifunction is in two aspects: by switching from metal to semiconducter phase, spin splitter is changed from spin splitter to $x\&y$ splitter; From low wave band to high wave band, the split orientation is inversed.
\par
By employing a spin-encoded scheme, we achieve NOT, BUF, OR, NAND logic gate and they can be switched from one to another by selecting different $\ce{VO2}$ phase and L/R output monitor.
\par
Finally, we design a periodical U-shaped nanostructures as a optical conveyor belt, so that using appropriate activation light with polarization phase difference constantly varying allows nanoparticles transporting. The two phases of $\ce{VO2}$ also make the direction of transporting controlable.

\section{Layout and Simulating Method}
\begin{figure}
\centering
\begin{tabular}{cc}
        \includegraphics[scale=0.15]{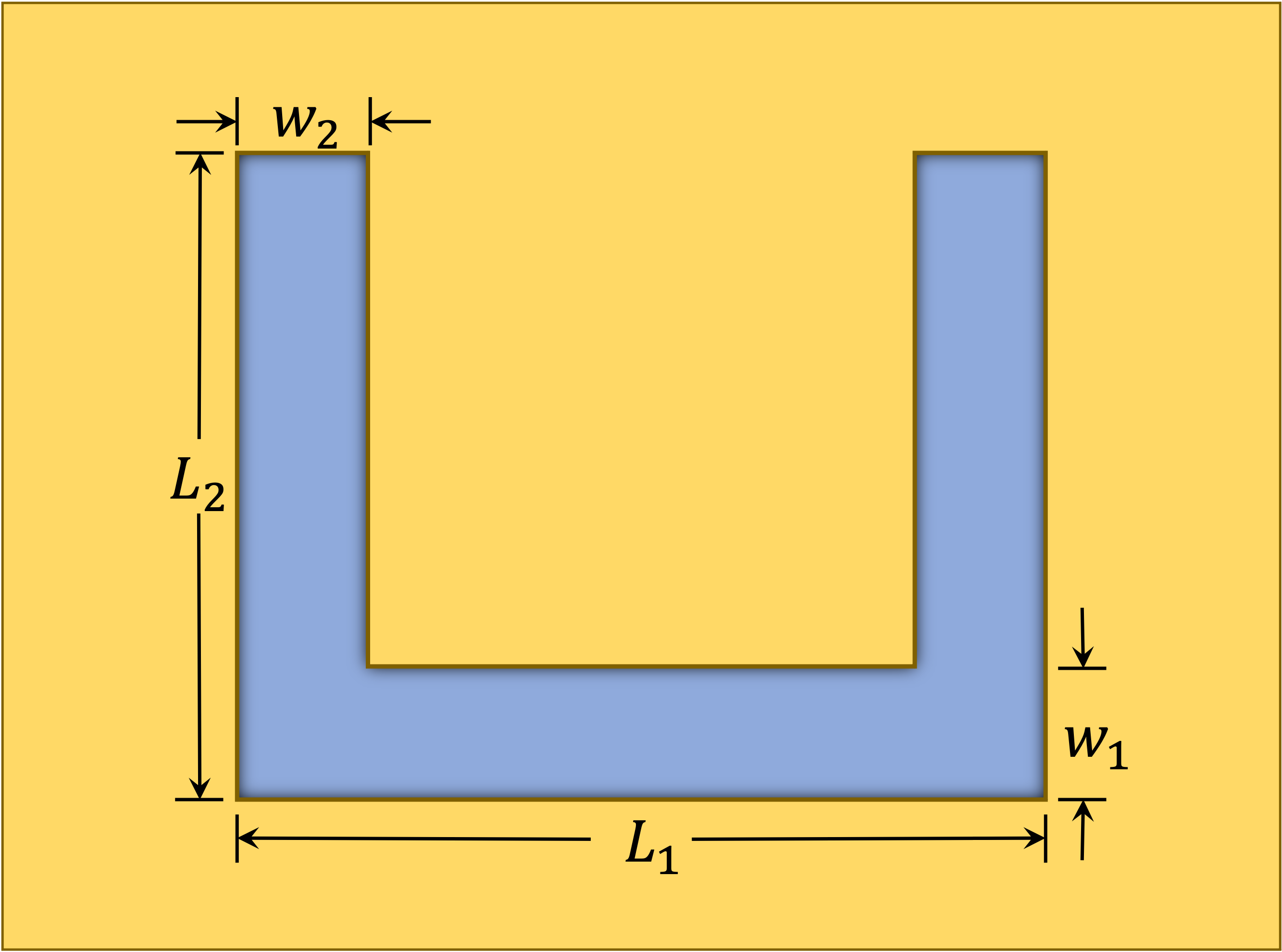} &
        \includegraphics[scale=0.2]{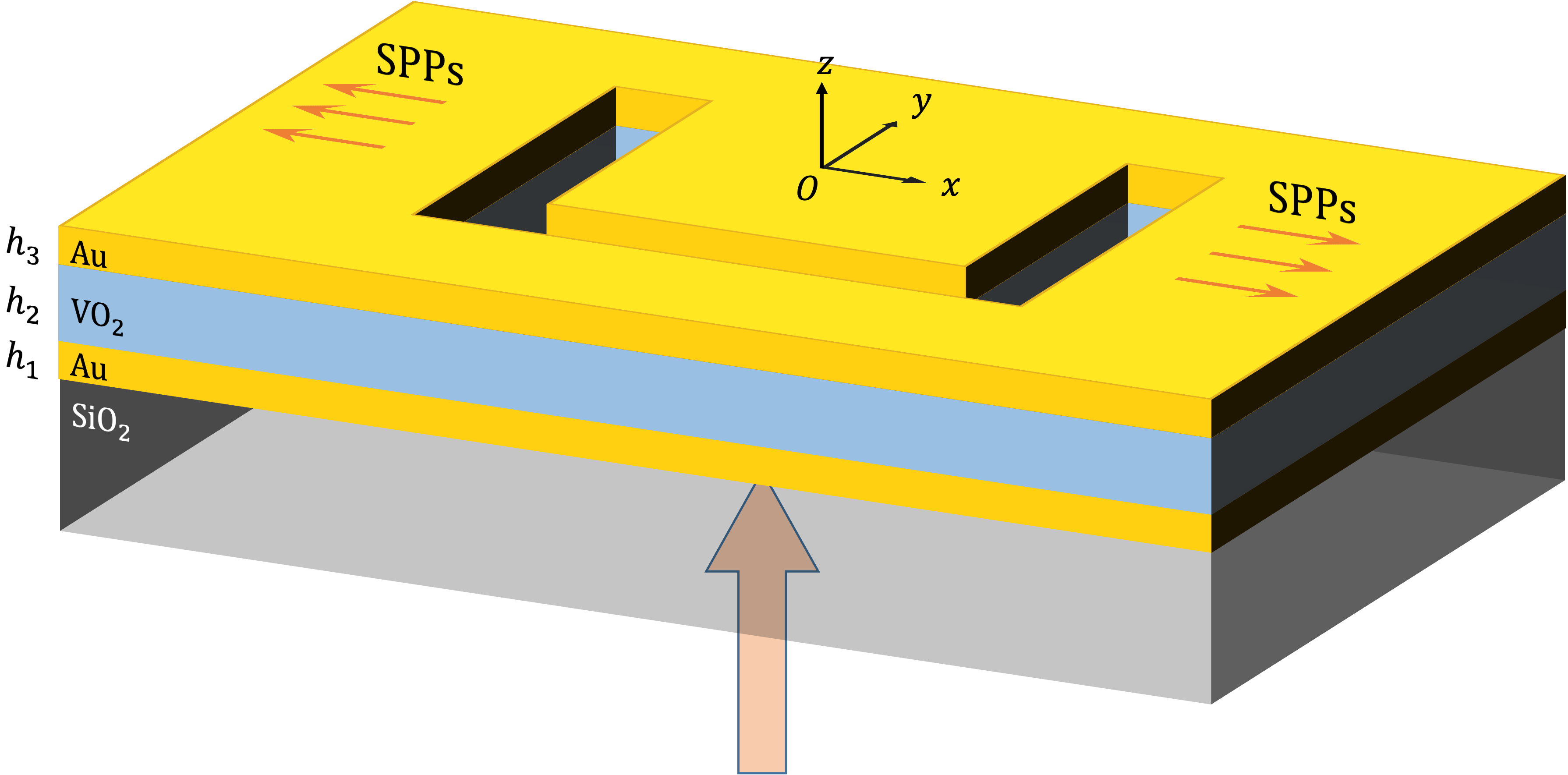}
    \end{tabular}

\caption{U-shaped hole's layout}\label{fig-Uhole-layout}
\end{figure}

As shown in Fig.\ref{fig-Uhole-layout} for the symmetrical U-shaped nanoholes, multifunctional nanostructure can be designed. By approximating the U-shaped hole as a combination of three rectangular holes, the SPP resonant mode can be analyzed qualitatively: SPP resonance in the left arm of U-shape is interfered with the bottom arm and right arm. The bottom arm is activated by $y$-polarization, and the left and right arms are activated by $x$-polarization. When the phase difference of interference is proper, the SPP amplitude in left arm can reach a maximum.
\par
\begin{figure}
\centering
\includegraphics[scale=0.32]{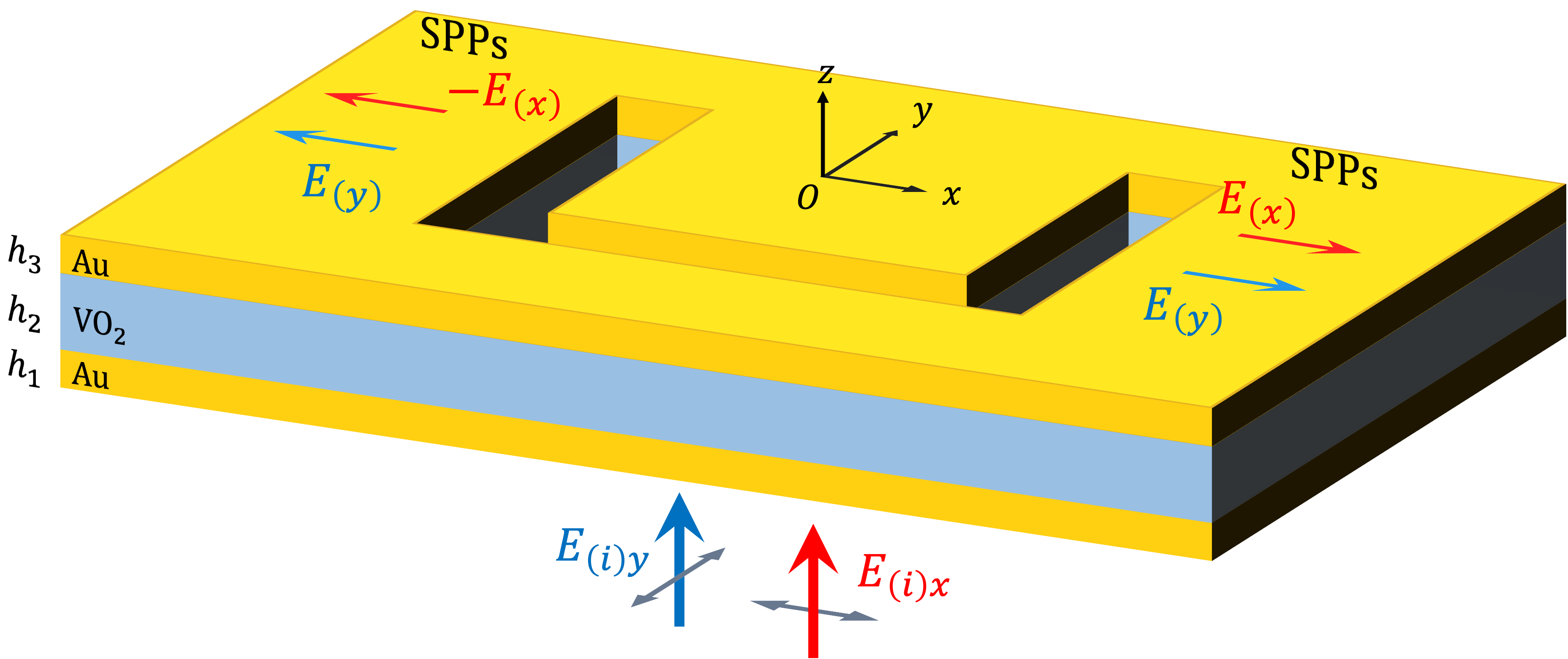}
\caption{U-hole's eigenstates of $x-,y-$polarization}
\label{fig-U-eigen}
\end{figure}

To analyze quantitively, the SPPs mode can be decomposed as a coherent superposition of two eigenmodes, i.e., the incident polarization state is a coherent superposition of the SPP eigenstate excited by $x-$ and $y-$polarization. The electric field of SPP activated by x-polarization is left-right antisymmetrically distruibuted, and the SPP activated by y-polarization is left-right symmetric, as shown in Fig.\ref{fig-U-eigen}. In the actual simulation, we can measure the SPP electric field intensity distribution of a single U-shaped hole above the uppermost gold film using $x-$ and $y-$ polarized light respectively, and adjust the parameters to get the resonance properties in different polarization states.
\par

\begin{figure}
\centering
\includegraphics[scale=0.25]{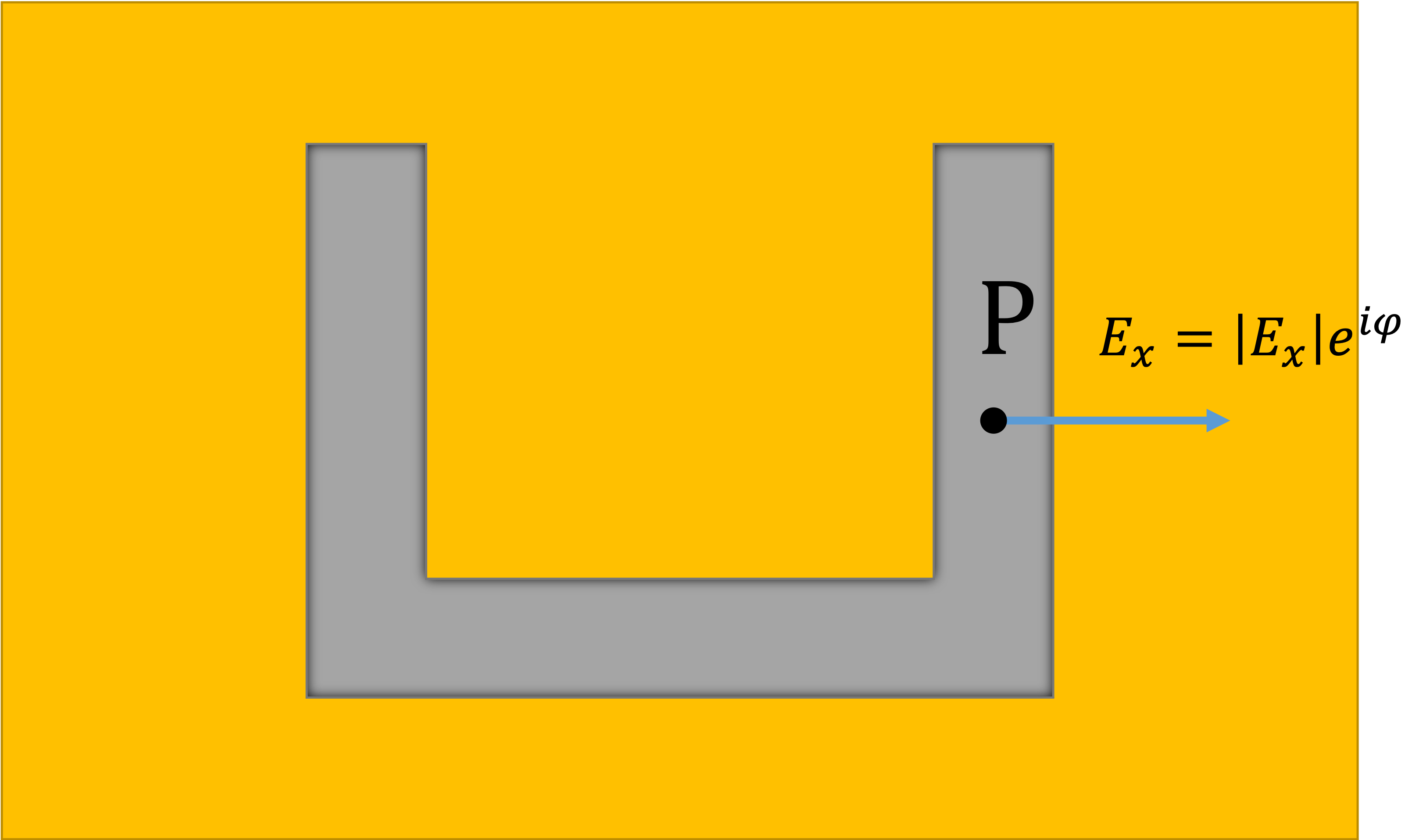}
\caption{Location of sampling point P}
\label{fig-U-Ppoint}
\end{figure}

A point monitor P (see Fig.\ref{fig-U-Ppoint}) at the center of the right arm can be used to simulate more efficiently, and its intensity and phase spectrum of $E_x$ in $x\&y$ eigenstates is recorded. Here the SPP local near-field distribution and nonlocal far-field distribution of U-hole are studied based on the P point's electric field data around telecom wavelength (1000-2400nm) in 2 eigenmodes.

\par
\paragraph{Near-field Distribution}
In the local area, we focus on the asymmetrical distribution of intensity on the left and right arms. It can be expected that only a single channel (e.g. the left channel) of the three will have a concentrated field distribution under appopriate parameters and polarization.
\par
With the incident light that has a phase difference $\sigma \pi/2$ between $E_x$ and $E_y$, the electric field amplitude $|A_{±P}|$ at P point and its point of symmetry with respect to the $y$-axis (denoted as -P) can be expressed as
\[|A_{\pm \text{P}}|^2=|A_{(x)}|^2 +|A_{(y)}|^2\pm 2|A_{(x)}||A_{(y)}|\cos \delta_{xy}\]
Where $A_{(x)}$ denotes the electric field in x-direction at P that excited by $E_x$, and $A_{(y)}$ denotes the electric field in x-direction at P that excited by $E_y$. The phase difference $\delta_{xy}$ of $A_{(x)}$ and $A_{(y)}$ is composed of two parts, one is the phase difference $\sigma \pi/2$ between the two directions of the electric field in the incident polarization state, which represents $x'-$linearly polarization or LCR respectively when $\sigma=0$ or $1$; The other part is the intrinsic phase difference between the two SPP modes (i.e., the phase difference when the incident polarization state is $x'-$linearized), $\Delta \varphi_{xy}$. If the parameters meet the following conditions,
\begin{equation}
\gamma=\frac{|A_{(x)}|}{|A_{(y)}|}=1,\delta_{xy}=\sigma \frac{\pi}{2}+\Delta \varphi_{xy}=0 \ \text{or} \ \pi
\label{eq-line-split} 
\end{equation}
Then the U-hole has a significant directional response to the incident wave, with the SPP being enhanced at the P point (or -P) in the right channel and the interference eliminating in the opposite direction. For the orthogonal polarization state, the SPP is enhanced at the -P (or P) position, so that the anisotropic response to the linear and circular polarization states can be achieved.
\par
The normalized SPP response intensity in the ±P direction is
\begin{align}
& I_{\pm \text{P}}=\frac{{}1}{2}\left(1\pm \frac{2\gamma}{1+\gamma^2}\cos\delta_{xy}\right)
\label{eq-intensity}\\
& \delta_{xy}=
\begin{cases}
\Delta\varphi_{xy}, & x'-\text{linear state}\\
\Delta\varphi_{xy}+\pi/2, & \text{left circular state}
\end{cases}
\end{align}
In the above equation, the intensity $I_{±\mathrm{P}}$ is 1 when the interference is large and 0 when it is small, so the near-field properties at the left and right channels of the U-hole can be analyzed once the P point response data of the metallic and semiconductor phases in the $x-$ and $y-$eigenmodes are obtained.

\begin{figure}
\centering
\includegraphics[scale=0.3]{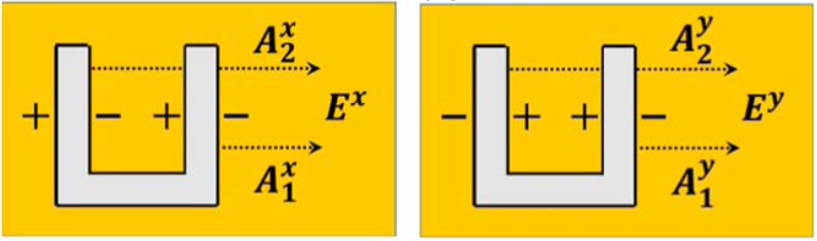}
\caption{The calculation of $x\&y$-eigenmode, figure from \cite{spin-control-Ushape}.}
\label{fig-farcal}
\end{figure}

\paragraph{Far-field Distribution}
When it comes to the far-field distribution, we are concerned with the preferred direction of the transmission to the far side of the U-hole response SPP, so we need to analyze the superposition of the electric field of different channels at the far field. As shown in the Fig.\ref{fig-farcal}, let the total span of the U-hole in $x$ direction be $d$, and the electric field in the $+x$ direction can be written as the superposition of that of two left and right arms as

\begin{equation}
\begin{aligned}
E_{(x)} (x&>\frac{d}{2}, y=0 )=A_{1(x)} \frac{e^{-\kappa(x-d / 2)}}{\sqrt{x-d / 2}} \\
& \times e^{i \varphi_{x}} e^{i k_{\text {spp }}(x-d / 2)}+A_{2(x)} \frac{e^{-\kappa(x+d / 2)}}{\sqrt{x+d / 2}} \\
& \times e^{i \varphi_{x}} e^{i k_{\text {spp }}(x+d / 2)}
\end{aligned}
\end{equation}

Due to the antisymmetry of the $x$ eigenmode excitation, the two amplitudes are equal and can be taken as the monitored P point amplitude, i.e. $A_{1(x)} = A_{2(x)} = A_{(x)}$. $\varphi_x$ is its phase, and the fractional term is the decay term, which can be neglected since the two channels' decay are equal at the far field; therefore, it can be abbreviated as

\begin{equation}
E_{(x)}(+x) \approx 2 A_{(x)}  \cos \frac{k_\text{spp}d}{2} e^{i\left(k_{\text{spp}} x+\varphi_{x}\right)}
\end{equation}

\par
For $y-$eigenmode, the two channel amplitudes take $A_{1(y)} = -A_{2(y)} = A_{(y)}$ due to antisymmetry, and it can be similarly derived that the SPP intensity at $+x$ is

\begin{equation}
E_{(y)}(+x) \approx 2 A_{(y)}  \sin \frac{k_\text{spp}d}{2} e^{i\left(k_{\text{spp}} x+\varphi_{y}-\frac{\pi}{2}\right)}
\end{equation}

Let the phase difference between the $x$ adn $y$ components of the electric field in the incident polarization state be $\sigma \pi/2$, then the SPP intensity obtained by the interfering superposition of the two state at the far field is 

\begin{equation}
\begin{aligned}
&|E(+x)|^{2}=4\left[\left(A_{(x)} \cos \frac{k_{\mathrm{spp}} d}{2}\right)^{2}+\left(A_{(y)} \sin \frac{k_{\mathrm{spp}} d}{2}\right)^{2}\right. \\
&\left.\quad+2 A_{(x)} A_{(y)} \cos \frac{k_{\mathrm{spp}} d}{2} \sin \frac{k_{\mathrm{spp}} d}{2} \cos \left(\Delta \varphi_{x y}+(\sigma-1) \frac{\pi}{2}\right)\right]
\end{aligned}
\end{equation}

\par
To make the SPP response in the far field significantly directional, it is required that

\begin{equation}
\gamma \cot \left(\frac{k_{\text{spp}}d}{2}\right)=1,\ \Delta \varphi_{xy}+(\sigma-1)\frac{\pi}{2}=0\  \text{or}\  \pi
\end{equation}

Then the U-hole structure has a significant directional response to the incident wave marked with $\sigma$, where the SPP is enhanced at $+x$ and the interference is eliminated in the opposite direction. For the orthogonal polarization state, the SPP is enhanced at the $-x$ position, so that the beam splitting between the linear and circular polarization states can be achieved.
\par
The SPP intensity in the $\pm x$ direction after normalization is
\begin{equation}
\begin{aligned}
I(\pm x)&=\frac{1}{2} \left[1\pm \frac{2\gamma \cot(k_{\text{spp}}d/2)}{1+\gamma^2 \cot^2(k_{\text{spp}}d/2)}\cos \delta_{xy}\right]
\\
& \delta_{xy}=\Delta \varphi_{xy}+(\sigma-1)\frac{\pi}{2}
\end{aligned}
\label{eq-U-far-I}
\end{equation}

\begin{figure}
\centering
\includegraphics[scale=0.32]{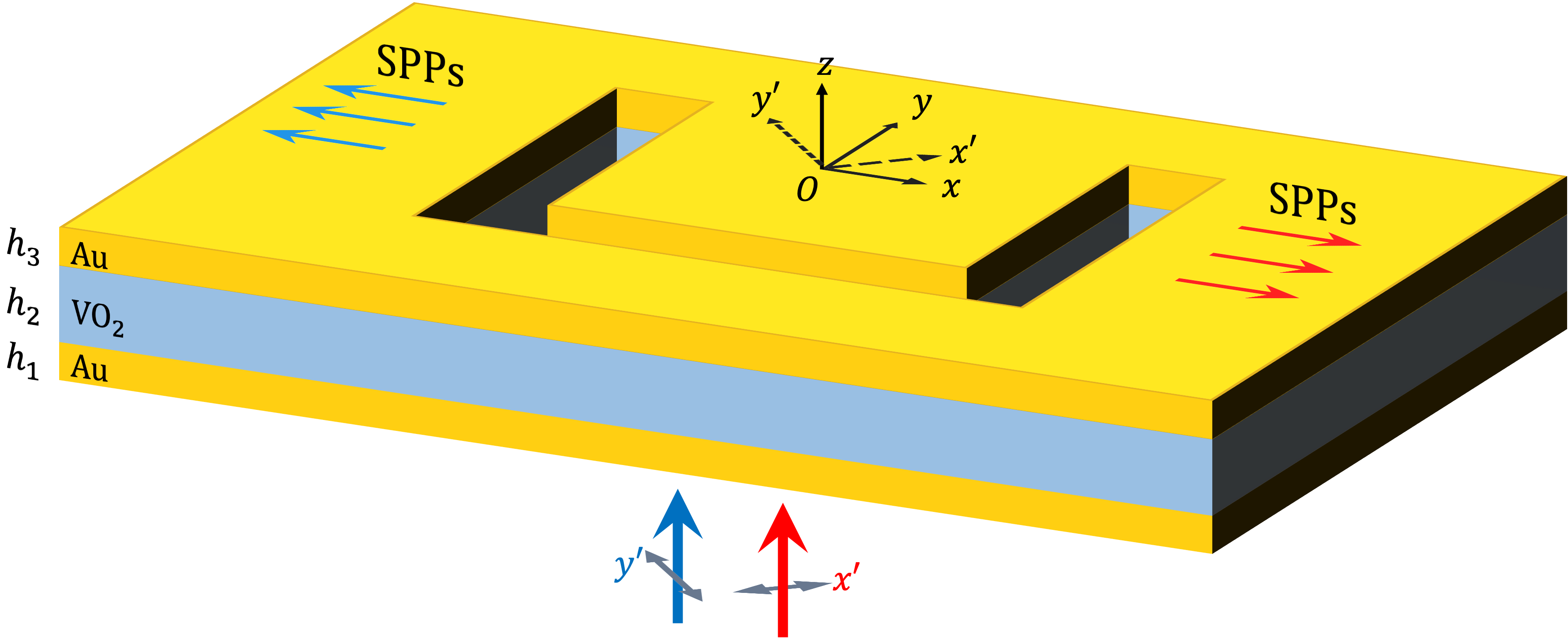}
\caption{SPP splitting in linear polarization}
\label{fig-U-line-split}
\end{figure}

\begin{figure}
\centering
\includegraphics[scale=0.32]{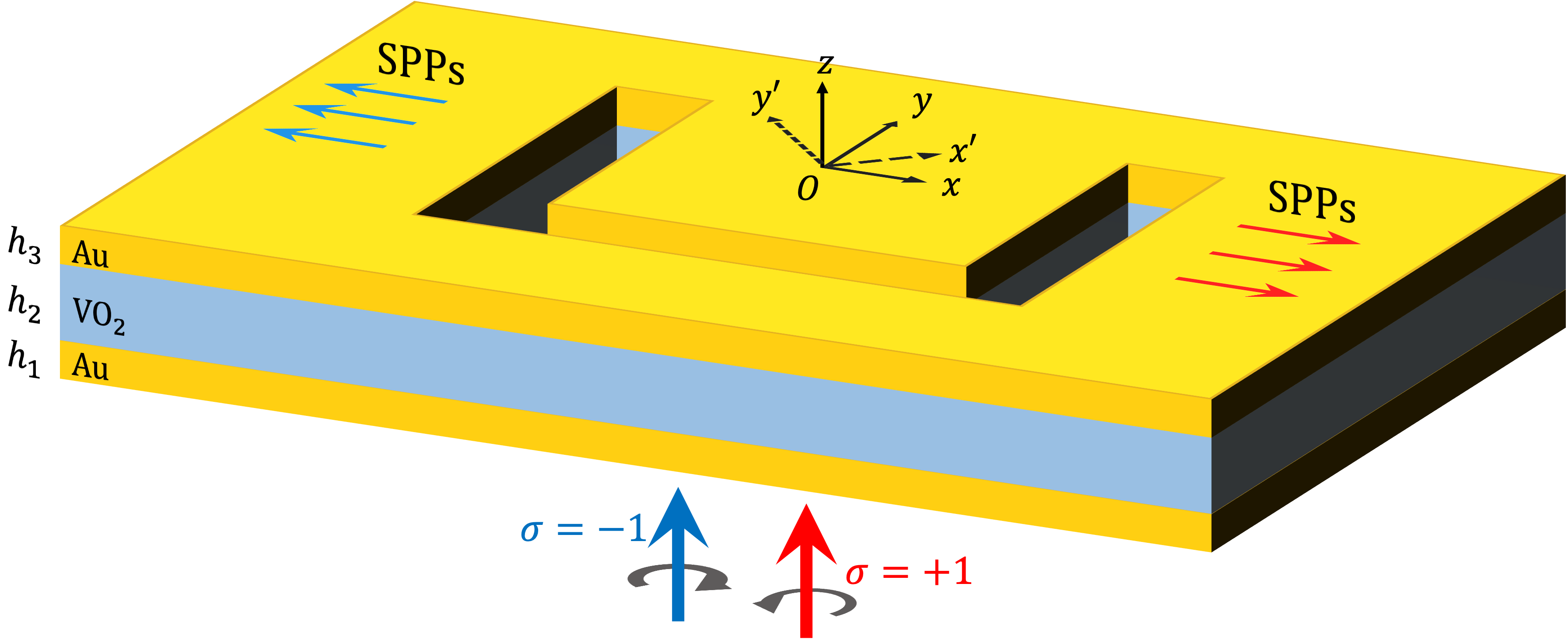}
\caption{SPP splitting in circular polarization}
\label{fig-U-circ-split}
\end{figure}

$\sigma=0,2$ is the line polarization state, and the beam splitting is shown in Fig.\ref{fig-U-line-split}; $\sigma=1,-1$ is the circular polarization state, and the beam splitting is shown in Fig.\ref{fig-U-circ-split}.

\section{Polarization Splitter}

We found in our FDTD simulations that a good beam splitting effect can be achieved when the SPP response of the far field is directional(see \ref{eq-U-far-I}) using a single vertical array of U-shaped holes. In the low waveband, the U-array becomes respectively the LCP/RCP splitter anbd the x-/y- linearized light splitter in $\ce{VO2}$'s metallic and semiconductive phase. In higher band, U-array is still the splitter of the same polarization in both phase, but the splitting direction of two orthogonal polarization state is exactly reversed.

\par
\begin{figure}
\centering
\includegraphics[scale=0.3]{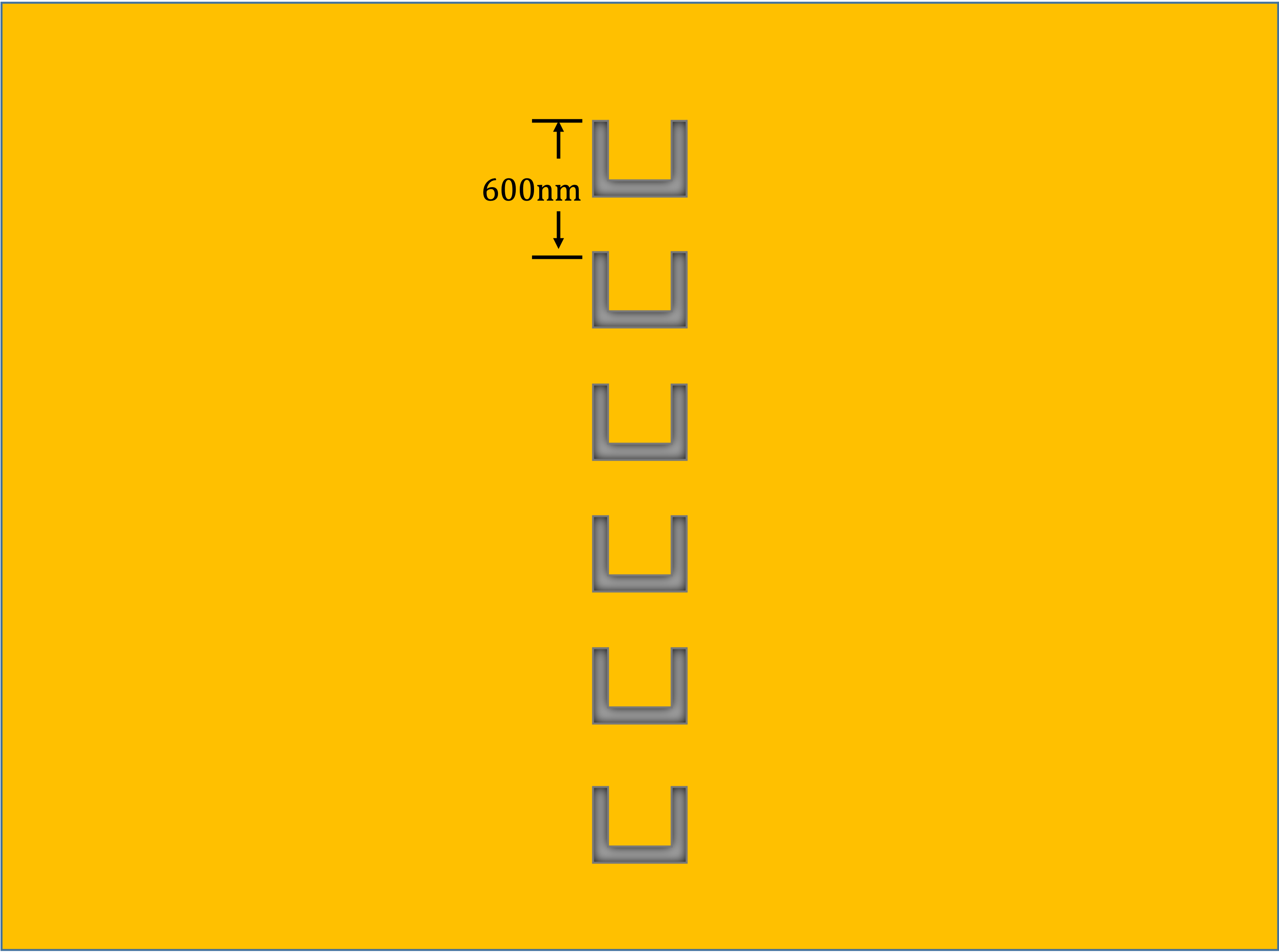}
\caption{The layout of U-array structure}
\label{U-array}
\end{figure}
\begin{figure}
\centering
\includegraphics[scale=0.42]{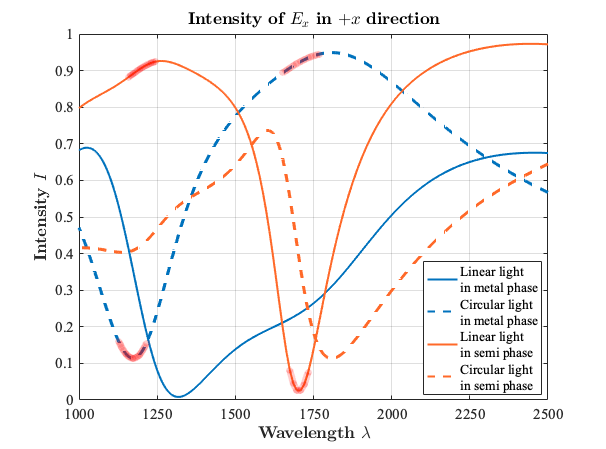}
\caption{The far-field intensity $I_{+x}$ of U-array's unit structure in $+x$ direction at different phases and different incident polarization states calculated from the data at point P; region marked in red is two bands that can realize splitting, where the far-field intensity in the $+x$ direction is close to 0\% or 100\%.}
\label{Array-intensity}
\end{figure}

The U-array is composed of several identical U-holes(Fig.\ref{U-array}) in a vertical column, and the parameter of each U-hole is $h_1=h_3=40\text{nm},\ h_2=470\si{nm},\ L_1=450\text{nm},\ w_1=70\text{nm},\ L_2=300\text{nm},\ w_2=50\text{nm}$. Using Eq.\ref{eq-U-far-I}, the far-field intensity of the two eigenstates measured in the 1000-2500nm waveband for a single structure is shown in Fig.\ref{Array-intensity}. 

\begin{figure}
\includegraphics[scale=0.3]{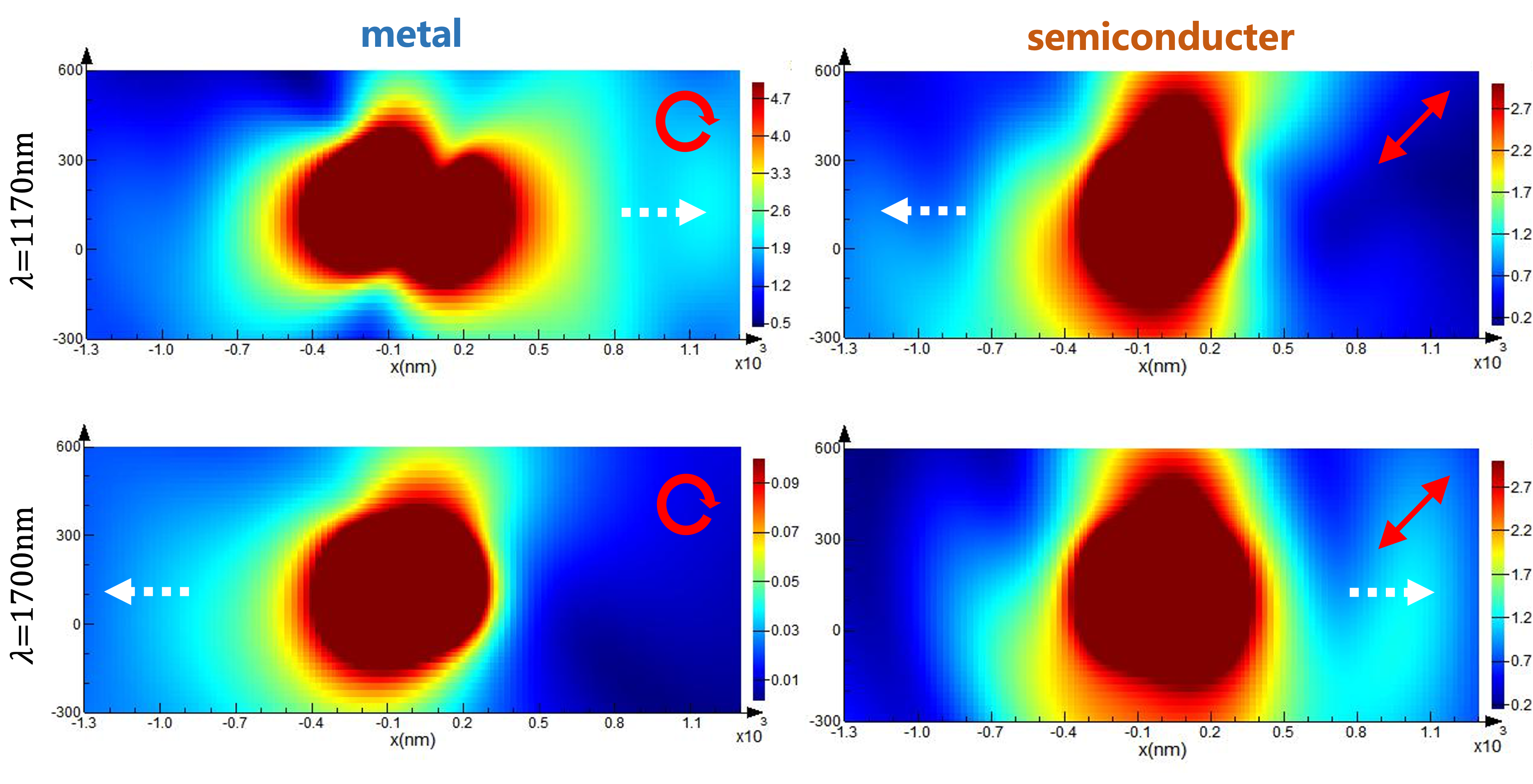}
\caption{The electric field distribution of U-array unit structure is plotted with the response of $x'-$polarization and LCP at low and high wavelengths in the metallic and semiconductor phases, respectively.}
\label{UII-single}
\end{figure}

\par
It can be seen that in the far-field distribution of U-array, the peaks and valleys of the circularly polarized incident's response in the metallic phase (blue dashed line) and the linearly polarized incident's response in the semiconductor phase (orange solid line) roughly coincide, and in the 1150-1200nm band, U-array in the metallic phase can split LCP to the $-x$ direction in the far-field, and U-array in the semiconductor phase can split $x’-$polarized light to the $+x$ direction. In the higher 1650-1750nm band, U-array in the metallic phase can split LCP to the $+x$ direction in the far-field, and U-array in the semiconductor phase can split $x’-$polarized light to the $-x$ direction. These theoretically enable multi-band and multi-polarization beam splitting with inversion. The electric field distribution measurement of the unit structure is shown in Fig.\ref{UII-single}.

\par
It can be seen that the unit structure reflects the difference of SPPs in two directions at the farther fields, but a single structure is not sufficient to realize splitting of polarization states. It is worth noting that the inversion of splitting direction requires a 180° phase change of the two SPP eigenstates at two different wavebands, which is a stringent condition and also results in a larger span of the two bands. If a square array is used, a suitable array period cannot be found that making the SPP field under both bands enhanced \footnote{Calculation shows that if a transverse period length of 1400 nm is taken, the intensity in the direction of splitting on both bands will be weakened by nearly 40\% compared to when wavelength is the period, and there is almost no splitting phenomenon. The working bandwidth of a square array is around 300-400nm.}.

\par
\begin{figure}
\centering
\includegraphics[scale=0.6]{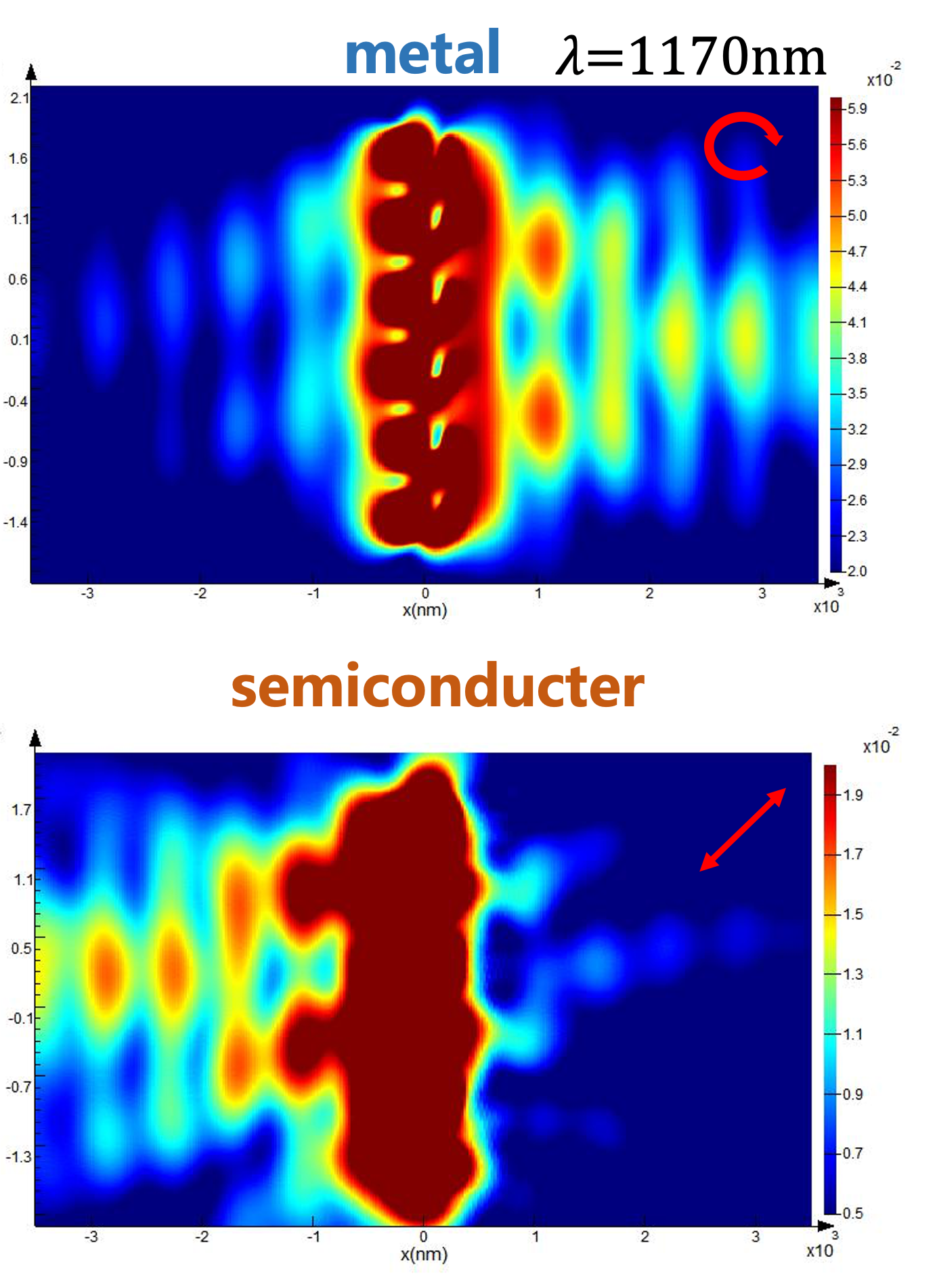}
\caption{The multifunction of U-array in low waveband}
\label{UII-array-1170}
\end{figure}
\begin{figure}
\centering
\includegraphics[scale=0.6]{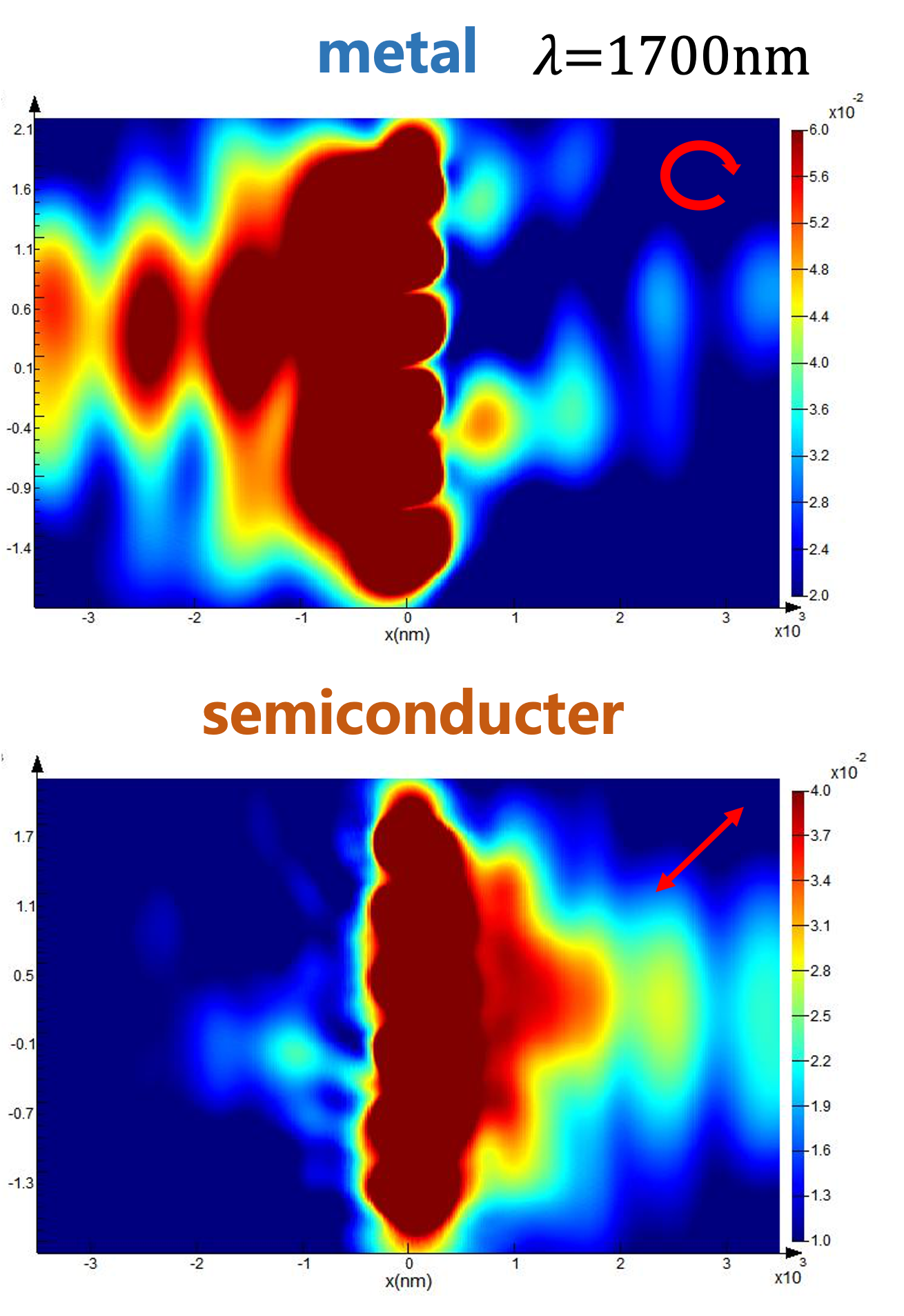}
\caption{The multifunction of U-array in high waveband}
\label{UII-array-1700}
\end{figure}

We therefore used an antenna like vertical array (see \cite{row-array}) with adjacent cells spaced at 600nm, as shown in Fig.\ref{U-array}. The responses of U-array for different material phases, different polarization states, and different wavelength bands are shown in Fig.\ref{UII-array-1170} and Fig.\ref{UII-array-1700}. It can be seen that the array already has excellent splitting property of dual band and multifunction with good compatibility for both bands in the case of using 6 cell structure to form the array, and it can be seen from the far-field intensity distribution Fig.\ref{Array-intensity} that there may be more bands with similar properties; therefore, based on U-array, it is possible to design multifunctional splitter for multiple bands, and also can form a circle-shaped periodic structure to realize the focus of SPP\cite{row-array}.

\section{Optical Logic Gates}

Optical logic gates are important elements in optical computing and optical circuits\cite{spin-encode}, but miniaturization of the logic gates to nanometer scale and functional switching of gates remain challenging. Using the SPP distribution of our U-hole at the far field, we propose a switchable optical logic gate with double-ended inputs and double-ended outputs in both \ce{VO2} phase. 

\par
The input polarization of optical logic gates can be decomposed into LCP and RCP, and here we use 1 and 0 to denote the two components. We stipulate that the spin state basis are $\boldsymbol{e}_1=(\hat{x}+\mathrm{i}\hat{y})/{\sqrt{2}}$ and $\boldsymbol{e}_0=(\mathrm{i}\hat{x}+\hat{y})/{\sqrt{2}}$. The double-ended inputs ``ab'' can be ``11'', ``00'', ``01'', ``10'', and they can be written uniformly as $\boldsymbol{e}_\mathrm{a}+\mathrm{i}\boldsymbol{e}_\mathrm{b}$. In this way ``01'' represents the x-polarization state, and ``10'' is the y-polarization state. 

\par
\begin{figure}
\centering
\includegraphics[scale=0.3]{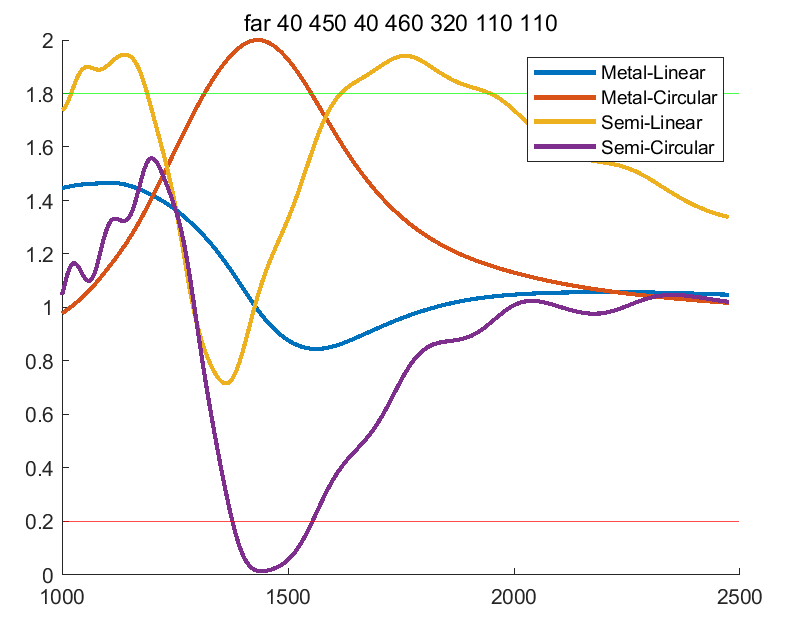}
\caption{The far-field intensity $I_{+x}$ of U-gate's structure at different phases and different incident polarization states calculated from the data at point P.}
\label{U-gate-intensity}
\end{figure}

Having two output ends (namd A and B for the left and right end) and two \ce{VO2} phases, we expect to have 4 different logic gates, which preferably should have a universal gate. After simulation, we choose parameter of the U-nanohole (the structure named U-gate in the following) to be $h_1=h_3=40\text{nm},\ h_2=450\si{nm},\ L_1=460\text{nm},\ w_1=110\text{nm},\ L_2=320\text{nm},\ w_2=110\text{nm}$, and the far-field intensity based on date of P-point is shown in Fig.\ref{U-gate-intensity}.

\par
\begin{figure}
\centering
\includegraphics[scale=0.32]{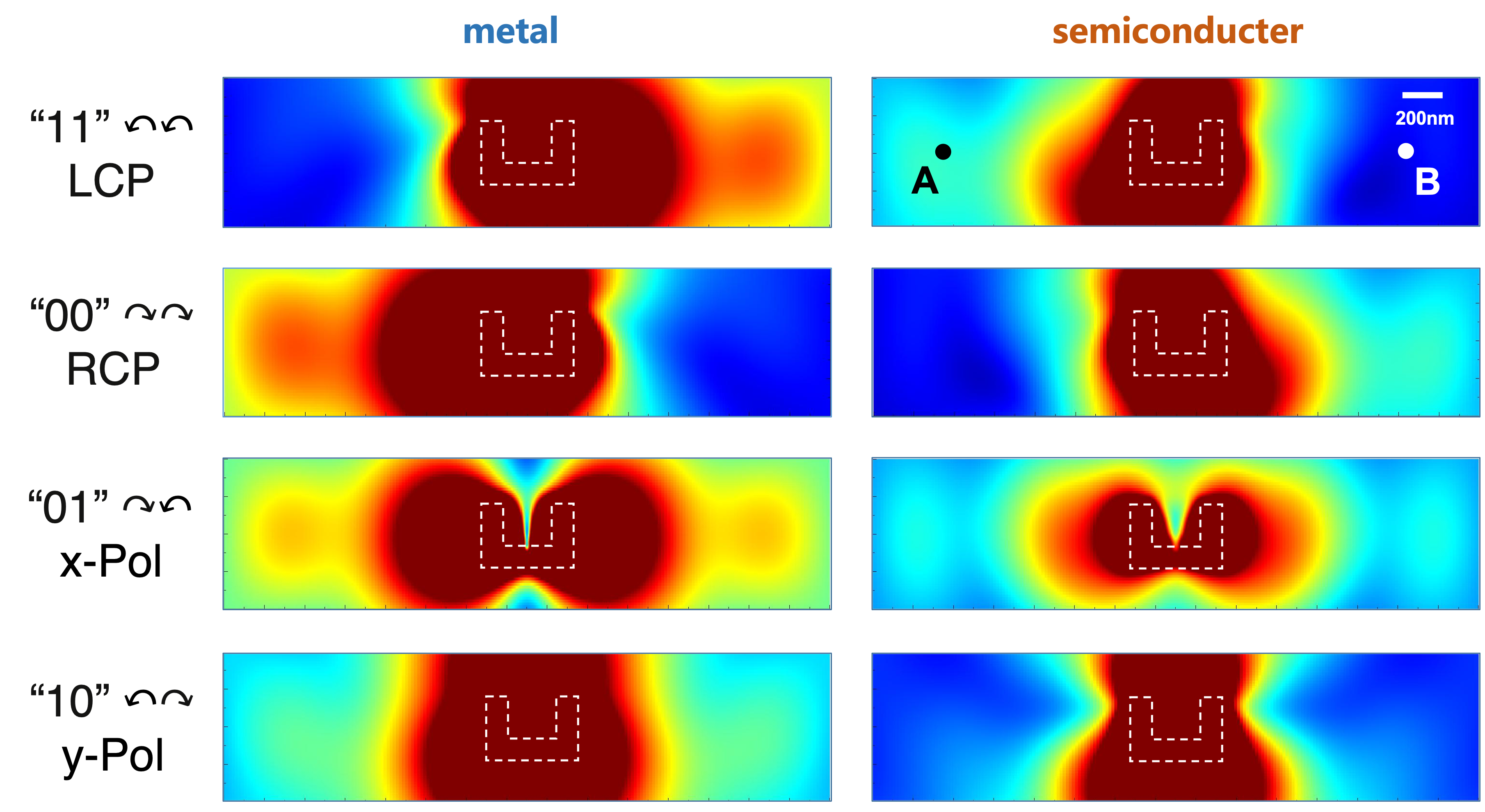}
\caption{The output field of U-gate with each input at $\lambda=1450 \si{nm}$, and the monitor plane is 20nm above the surface. A and B are two output monitor points.}
\label{U-gate-field}
\end{figure}

From Fig.\ref{U-gate-intensity} we can see that when LCP is the incident state, the far-field intensity at +x direction is maximal (using output signal ``1'' to denote it) in metallic phase and mininal (output signal ``0'') in semiconductive phase at wavelength of about 1450nm. As a result, the double-ended output signal ``AB'' is respectively ``01'' ``10'' in metallic phase with input ``11'' ``00'', and ``10'' ``01'' in semiconductive phase with the same input. This is a spin splitter with inversion at different \ce{VO2} phases, and with the calculation of field distribution in ``01'' and ``10'' states, we can give the output field distribution as shown in Fig.\ref{U-gate-field}.

\par
To get the output intensity quantitatively, we set the two monitor points A and B at coordinate $(\mp 1250\si{nm},0,20\si{nm})$, and they give the signal ``0'' or ``1'' according to a threshold. The output intensity and truth table of the U-gate is shown in Table \ref{truth-table}.

\begin{table}
\scriptsize
\begin{tabular}{|c|cc|cc|}
\hline
                    & \multicolumn{2}{c|}{\textbf{metallic}}   & \multicolumn{2}{c|}{\textbf{semiconductive}} \\ \hline
\textbf{Input ab} & \multicolumn{1}{c|}{\textbf{output A}} & \textbf{output B} & \multicolumn{1}{c|}{\textbf{output A}} & \textbf{output B} \\ \hline
11, LCP             & \multicolumn{1}{c|}{0 (5.66)} & 1 (27.6) & \multicolumn{1}{c|}{1 (14.8)}   & 0 (4.74)   \\ \hline
00, RCP             & \multicolumn{1}{c|}{1 (27.6)} & 0 (5.66) & \multicolumn{1}{c|}{0 (4.74)}   & 1 (14.8)   \\ \hline
01, x-Pol           & \multicolumn{1}{c|}{1 (23.4)} & 1 (23.4) & \multicolumn{1}{c|}{1 (13.8)}   & 1 (13.8)   \\ \hline
10, y-Pol           & \multicolumn{1}{c|}{1 (15.9)} & 1 (15.9) & \multicolumn{1}{c|}{0 (7.01)}   & 0 (7.01)   \\ \hline
\textbf{Expression} & \multicolumn{1}{c|}{$\mathrm{A=\overline{ab}}$}        & $\mathrm{B=a+b}$       & \multicolumn{1}{c|}{$\mathrm{A=b}$ }          & $\mathrm{B=\overline{a}}$           \\ \hline
\textbf{Gate}       & \multicolumn{1}{c|}{NAND}     & OR       & \multicolumn{1}{c|}{BUF}        & NOT        \\ \hline
\end{tabular}
\caption{The output amplitude of SPP field in point A and B (setting the amplitude of incident as 1, and $10^{-3}$ as  the unit of output amplitude), the corresponding truth table, the expressions and types of U-gate. The threshold amplitude is chosen as $10\times 10^{-3}$.}
\label{truth-table}
\end{table}

The output amplitudes in Table \ref{truth-table} correspond with those small maps in Fig.\ref{U-gate-field}. We specify that the signal is 0 for outputs with amplitudes less than 0.01 and 1 for outputs with amplitudes greater than 0.01. 

\par
In metallic phase, for input ``11'' and ``00'' the U-gate acts as a splitter so the output AB is ``01'' or ``10'', and the intensity ratio $R=|\boldsymbol{E}^{(\mathrm{B})}_{00}|^2/|\boldsymbol{E}^{(\mathrm{B})}_{11}|^2=|\boldsymbol{E}^{(\mathrm{A})}_{11}|^2/|\boldsymbol{E}^{(\mathrm{A})}_{00}|^2$ reaches a high value of 23.8. For input ``01'', the two arms of U-gate along the y-direction is excited by x-polarized electric field, so the output gives ``11''. For input ``10'' that is y-polarized, the arm along the x-direction is excited and the other two arms get excited indirectly by SPPs from the horizontal arm. In the metallic phase, the SPP's attenuation is smaller, so the SPP emitted from the vertical arm after its excitation has a large amplitude at points A and B, giving a logical signal of 1. 

\par
In semiconductive phase, for input ``11'' and ``00'' the U-gate acts as an inversed splitter compared to that in metallic phase, so the output AB is ``10'' or ``01'', and the intensity ratio $R=|\boldsymbol{E}^{(\mathrm{A})}_{00}|^2/|\boldsymbol{E}^{(\mathrm{A})}_{11}|^2=|\boldsymbol{E}^{(\mathrm{B})}_{11}|^2/|\boldsymbol{E}^{(\mathrm{B})}_{00}|^2$ reaches 9.8. For input ``01'' that is x-polarized, the output signal of A and B is still ``11'' due to the direct excitation; for input ``10'' that is y-polarized, the output signal is ``00''. This is because in the semiconductive phase, the SPP decays fast, and the indirect excitation is insignificant. 

\par
Table \ref{truth-table} gives the multifunction of the double-output U-gate. In semiconductive phase, the gates are respectively BUF and NOT gates of input ends b and a, which are relatively trivial, and they can be considered as two independent gates with signle-ended input and output. In the metallic phase, the two gates are not independent any more, and they are NAND and OR gates of the two input ends. It is worth mentioning that the NAND gate is a universal gate, which can implement any Boolean function without the need to use any other type of logic gate. Thus the U-gate is nontrivial in metallic phase compared to that in semiconductive phase.

\par
In contrast to some existing studies on nano-optical logic gates, our proposed U-gate has the following characteristics: we can control the type of the logic gates by changing the phase of \ce{VO2}, and this feature has never been used to regulate the logic gates before. This allows for a temperature-controlled optical logic circuit, like an optical switch. The double-ended input and output is also an innovation, which gives 4 possible logic gates. In addition, unlike other studies\cite{spin-encode}, our input signals ``01'' and ``10'' are defined to be different, and we can easily distinguish between them. In other words, input ends A and B are not interchangeable; they are distinguishable in semiconductive phase.

\par
The input singals ``01'' and ``10'' can also be defined to be the same. We can define $\boldsymbol{e}_1=(\hat{x}+\mathrm{i}\hat{y})/\sqrt{2}$, $\boldsymbol{e}_0=(\hat{x}-\mathrm{i}\hat{y})/\sqrt{2}$, and ``ab'' is $(\boldsymbol{e}_{\mathrm{a}}+\boldsymbol{e}_{\mathrm{b}})/\sqrt{2}$. Thus ``01'' and ``10'' both refer to the x-polarized state, and we name this definition X. Similarly, we can define $\boldsymbol{e}_1=(\hat{y}-\mathrm{i}\hat{x})/\sqrt{2}$, $\boldsymbol{e}_0=(\hat{y}+\mathrm{i}\hat{x})/\sqrt{2}$, and ``ab'' is $(\boldsymbol{e}_{\mathrm{a}}+\boldsymbol{e}_{\mathrm{b}})/\sqrt{2}$, so ``01'' and ``10'' refer to the y-polarized state. We name this definition Y. The output table in these two definitions is shown in Table \ref{truth-table-XY}. We can see that two new gates are available, including a universal gate NOR, and we can construct six types of gate in total.

\begin{table}
\scriptsize
\begin{tabular}{|c|cc|cc|}
\hline
       Definition X             & \multicolumn{2}{c|}{\textbf{metallic}}   & \multicolumn{2}{c|}{\textbf{semiconductive}} \\ \hline
\textbf{Input ab} & \multicolumn{1}{c|}{\textbf{output A}} & \textbf{output B} & \multicolumn{1}{c|}{\textbf{output A}} & \textbf{output B} \\ \hline
11, LCP             & \multicolumn{1}{c|}{0 (5.66)} & 1 (27.6) & \multicolumn{1}{c|}{1 (14.8)}   & 0 (4.74)   \\ \hline
00, RCP             & \multicolumn{1}{c|}{1 (27.6)} & 0 (5.66) & \multicolumn{1}{c|}{0 (4.74)}   & 1 (14.8)   \\ \hline
01, x-Pol           & \multicolumn{1}{c|}{1 (23.4)} & 1 (23.4) & \multicolumn{1}{c|}{1 (13.8)}   & 1 (13.8)   \\ \hline
10, x-Pol           & \multicolumn{1}{c|}{1 (23.4)} & 1 (23.4) & \multicolumn{1}{c|}{1 (13.8)}   & 1 (13.8)   \\ \hline
\textbf{Expression} & \multicolumn{1}{c|}{$\mathrm{A=\overline{ab}}$}        & $\mathrm{B=a+b}$       & \multicolumn{1}{c|}{$\mathrm{A=a+b}$ }          & $\mathrm{B=\overline{ab}}$           \\ \hline
\textbf{Gate}       & \multicolumn{1}{c|}{NAND}     & OR       & \multicolumn{1}{c|}{OR}        & NAND        \\ \hline
\multicolumn{5}{c}{}\\
\hline
       Definition Y             & \multicolumn{2}{c|}{\textbf{metallic}}   & \multicolumn{2}{c|}{\textbf{semiconductive}} \\ \hline
\textbf{Input ab} & \multicolumn{1}{c|}{\textbf{output A}} & \textbf{output B} & \multicolumn{1}{c|}{\textbf{output A}} & \textbf{output B} \\ \hline
11, LCP             & \multicolumn{1}{c|}{0 (5.66)} & 1 (27.6) & \multicolumn{1}{c|}{1 (14.8)}   & 0 (4.74)   \\ \hline
00, RCP             & \multicolumn{1}{c|}{1 (27.6)} & 0 (5.66) & \multicolumn{1}{c|}{0 (4.74)}   & 1 (14.8)   \\ \hline
01, y-Pol           & \multicolumn{1}{c|}{1 (15.9)} & 1 (15.9) & \multicolumn{1}{c|}{0 (7.01)}   & 0 (7.01)   \\ \hline
10, y-Pol           & \multicolumn{1}{c|}{1 (15.9)} & 1 (15.9) & \multicolumn{1}{c|}{0 (7.01)}   & 0 (7.01)   \\ \hline
\textbf{Expression} & \multicolumn{1}{c|}{$\mathrm{A=\overline{ab}}$}        & $\mathrm{B=a+b}$       & \multicolumn{1}{c|}{$\mathrm{A=ab}$ }          & $\mathrm{B=\overline{a+b}}$           \\ \hline
\textbf{Gate}       & \multicolumn{1}{c|}{NAND}     & OR       & \multicolumn{1}{c|}{AND}        & NOR        \\ \hline
\end{tabular}
\caption{The output amplitude, the corresponding truth table, the expressions and types of U-gate in definition X and Y. The threshold amplitude is chosen as $10\times 10^{-3}$.}
\label{truth-table-XY}
\end{table}

\section{Optical conveyor Belt}
The U-hole structure can also be used to transport nano-particles\cite{belt-1}\cite{belt-2}, i.e. the optical tweezers. With the strong near-field hot spots on the U-hole, an optical trap is produced, and by changing the polarization in a proper way, the trap moves continuously in one direction. Thus a nano-particle can be peristalticly transported. Based on the SPP distribution of our U-hole at the near field and the phase transition of \ce{VO2} film, we propose two types of switchable optical conveyor belt, on which we can realize inversible transport with polarization varying respectively from x-polarized to y-polarized for the first type, or from LCP to RCP for the second type. 

\par
According to Rayleigh scattering theory, the force on a particle in a light field can be calculated by considering the particle as an electric dipole. The force is composed of scattering force and gradient force, where the scattering force is proportional to the light intensity
$$
F_{s}=\frac{128 \pi^{5} a^{6} n_{m}}{3 \lambda^{4} c}\left(\frac{h^{2}-1}{h^{2}+2}\right)^{2} I_{0}
$$
Where $a$ is the radius of particle, $n_m$ is the refractive index of material, $h$ is the ratio of the refractive index of the particle to the refractive index of the material. The gradient force on the particle is proportional to the gradient of the light intensity
\begin{equation}
F_{g}=\frac{n_{m}^{2} a^{3}}{2}\left(\frac{h^{2}-1}{h^{2}+2}\right)^{2} \nabla I_{0}
\end{equation}
The diameter of particle $d\sim 400\si{nm}$, and wavelength $\lambda \sim 2000\si{nm}$, so we can neglect the scattering force for an approximation, and treat the negtive value of the intensity as the equivalent potential energy
\begin{equation}
U=-I_0
\end{equation}
As a result, the hot spots in the near-field distribution are potential wells for nano-particles. 

\par
In our simulation, the particle is a polystyrene sphere and it is in the water above the surface of the top \ce{Au} film. For a polystyrene sphere with radius of 200nm, the velocity of thermal motion in water corresponds to a Reynolds number $\mathrm{Re} \approx 2 × 10^{-3} \ll 1$. The sphere is subject to large viscous forces, and when the forces are unbalanced they reach their maximum velocity almost instantaneously and quickly reach steady state, and the state of motion of the sphere is almost independent of the state of the previous moment. Therefore we can assume that the particle is always in the bottom of the trap. 

\begin{figure}
\centering
\includegraphics[scale=0.4]{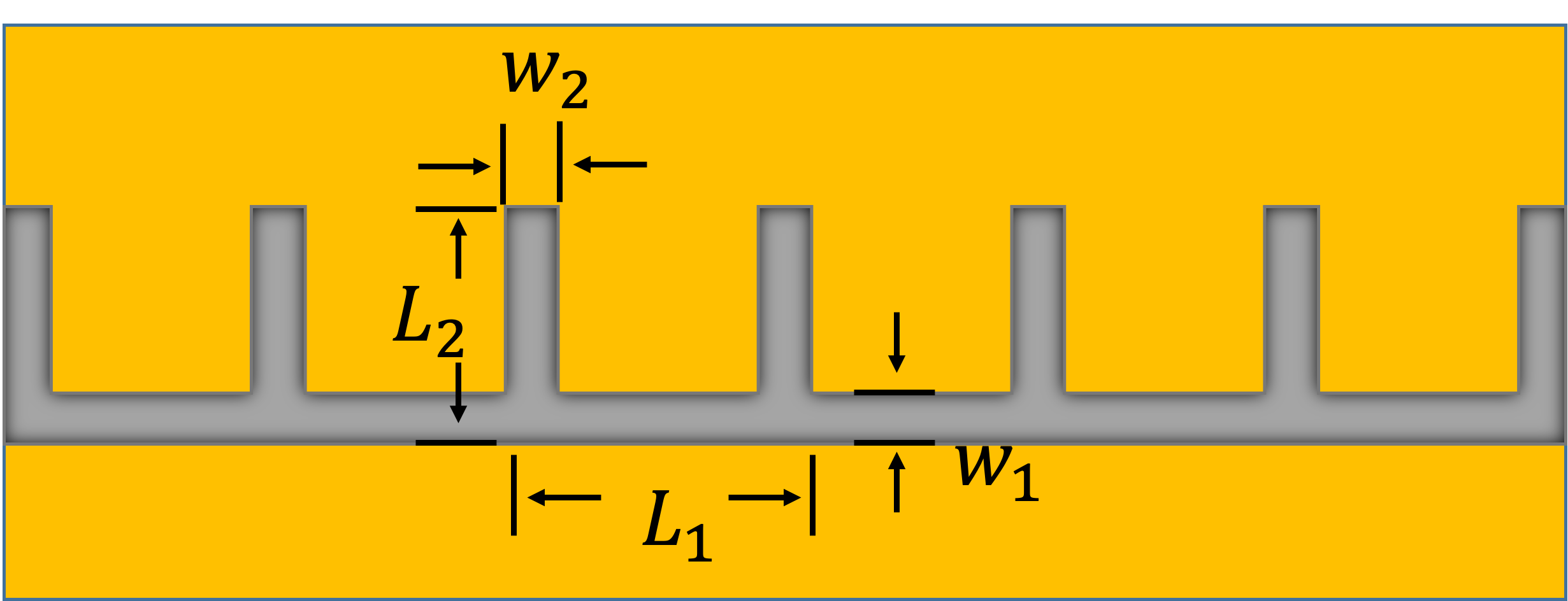}
\caption{The layout of periodic sturcture U-belt, and the medium in the nanohole and above the surface is water.}
\label{U-belt-layout}
\end{figure}

\par
We propose two types of the U-belt, and the first one (named UI-belt) is a periodical structure with parameter $h_1=h_3=40\si{nm},\ h_2=450\si{nm},\ L_1=560\si{nm},\ L_2=320\si{nm},\ w_1=80\si{nm},\ w_2=80\si{nm}$. The U-belt's layout is shown in Fig.\ref{U-belt-layout}, and some parameters are redefined. 

\par
\begin{figure}
\centering
\includegraphics[scale=0.3]{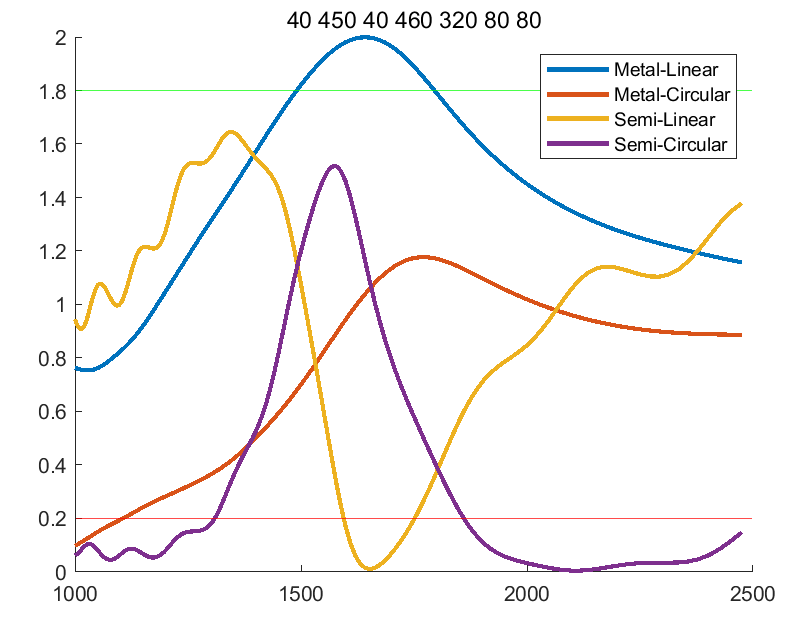}
\caption{The near-field intensity $I_{+P}$ of the prototype of UI-belt's unit structure at different phases and different incident polarization states. The real unit structure of UI-belt has its $L_1$ 100nm larger than the prototype, and two adjacent arms are merged into one.}
\label{UI-belt-single}
\end{figure}

 
The design of the UI-belt comes from the near-field intensity of the single U-hole with parameter $h_1=h_3=40\si{nm},\ h_2=450\si{nm},\ L_1=460\si{nm},\ L_2=320\si{nm},\ w_1=80\si{nm},\ w_2=80\si{nm}$ (only $L_1$ is different from UI-belt), as shown in Fig.\ref{UI-belt-single}. The prototype of UI-belt is a nearly perfect inversible splitter for x'- and y'-polarized states, i.e. in metallic phase, for LCP the SPP is focused on the right arm, and for RCP focused on the left arm; in semiconductive phase it's inversed. While in metallic phase, for x-polarized state, the vertical arms are excited; for y-polarized state, the horizontal arm is excited. Therefore if we use linearly polarized light with constantly rotating polarization direction as the incident state, the hot spot of field may have a continuous move from one vertical arm to another.

\par
\begin{figure}
\centering
\includegraphics[scale=0.35]{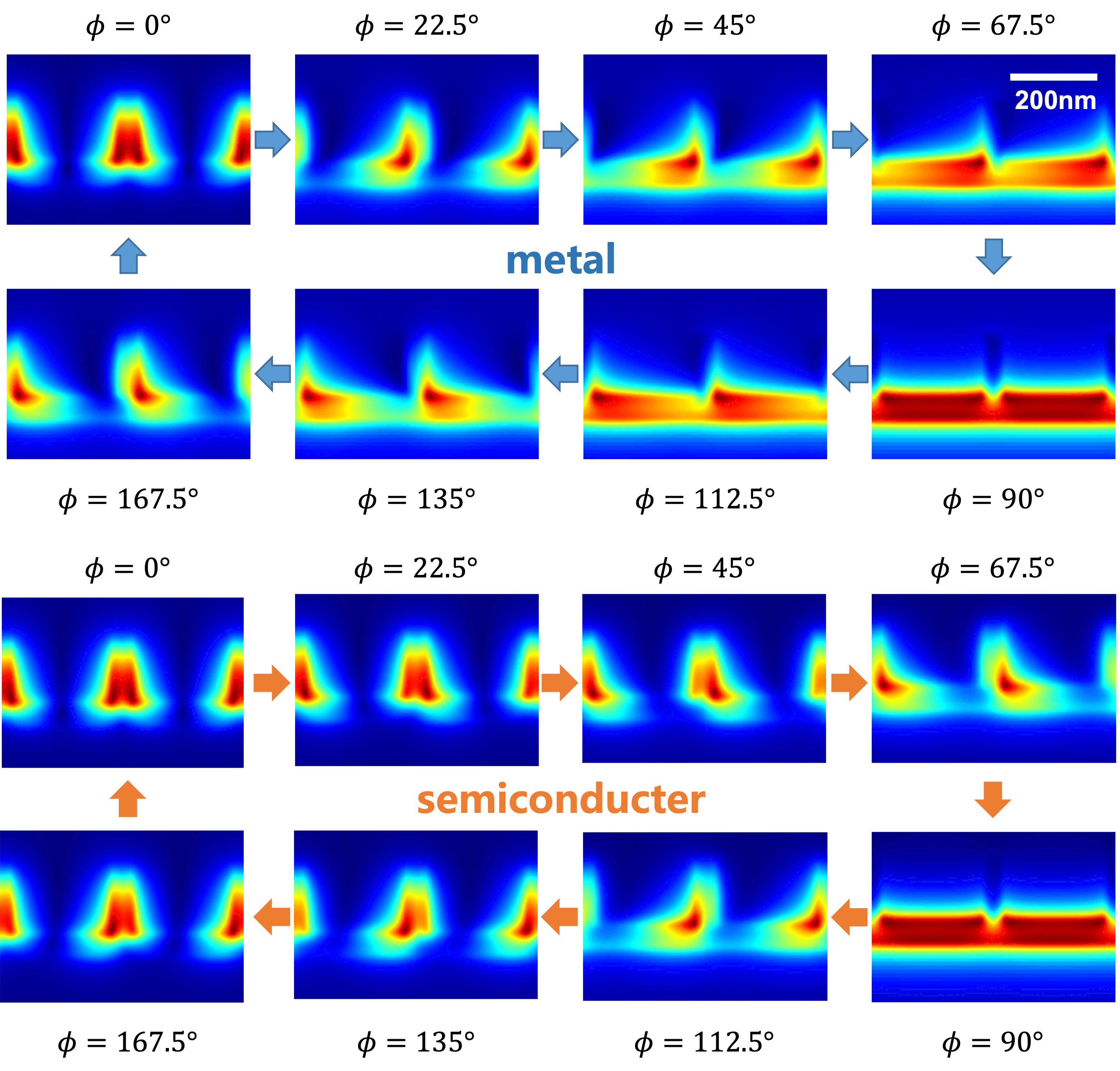}
\caption{The field distribution of UI-belt in two \ce{VO2} phases and different polarization directions, and $\lambda=2200\si{nm}$, $\phi$ is the angle between polarization direction and x-axis. The range of figure is two periods, and monitor plane is 20nm above the surface.}
\label{UI-belt-field}
\end{figure}

Combining these unit structures into periodic structures, the hot spot is expect to move from one unit to the next. The adjacent structures can affect the excitation of an individual structure, so the parameter needs to be adjusted. After sampling and analyzing we take $L_1\ 100\si{nm}$ larger, and the two adjacent arms are merged into one. The field distribution in rotating linear polarization is shown in Fig.\ref{UI-belt-field}.

\par
From Fig.\ref{UI-belt-field} we can see that when the polarization direction is rotating counterclockwise, the hot spots move from right to left in metallic phase and from left to right in semiconductive phase. In metallic phase it moves more continuously. By the way, the wavelength $\lambda$ should be multiplied by the index of water. 

\par
\begin{figure}
\centering
\includegraphics[scale=0.33]{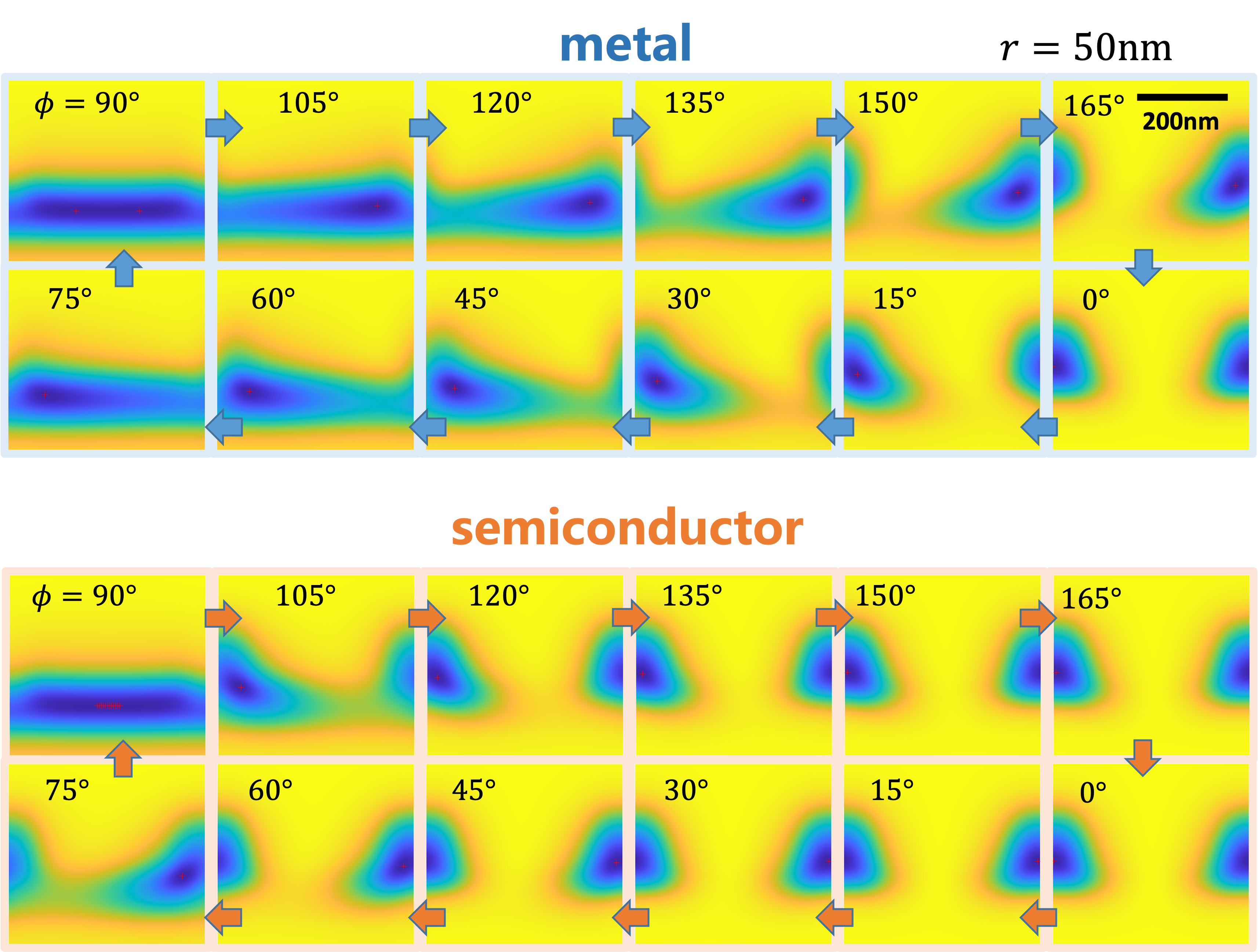}
\caption{The potential field of UI-belt for particle with $r=50\si{nm}$ in two \ce{VO2} phases and different polarization directions. The range of figure is one period, and the lowest point of the sphere is 20nm above the surface. The small red cross marks the location of the bottom of potential energy. Blue area means low energy and yellow denotes the high.}
\label{UI-belt-50}
\end{figure}
\begin{figure}
\centering
\includegraphics[scale=0.33]{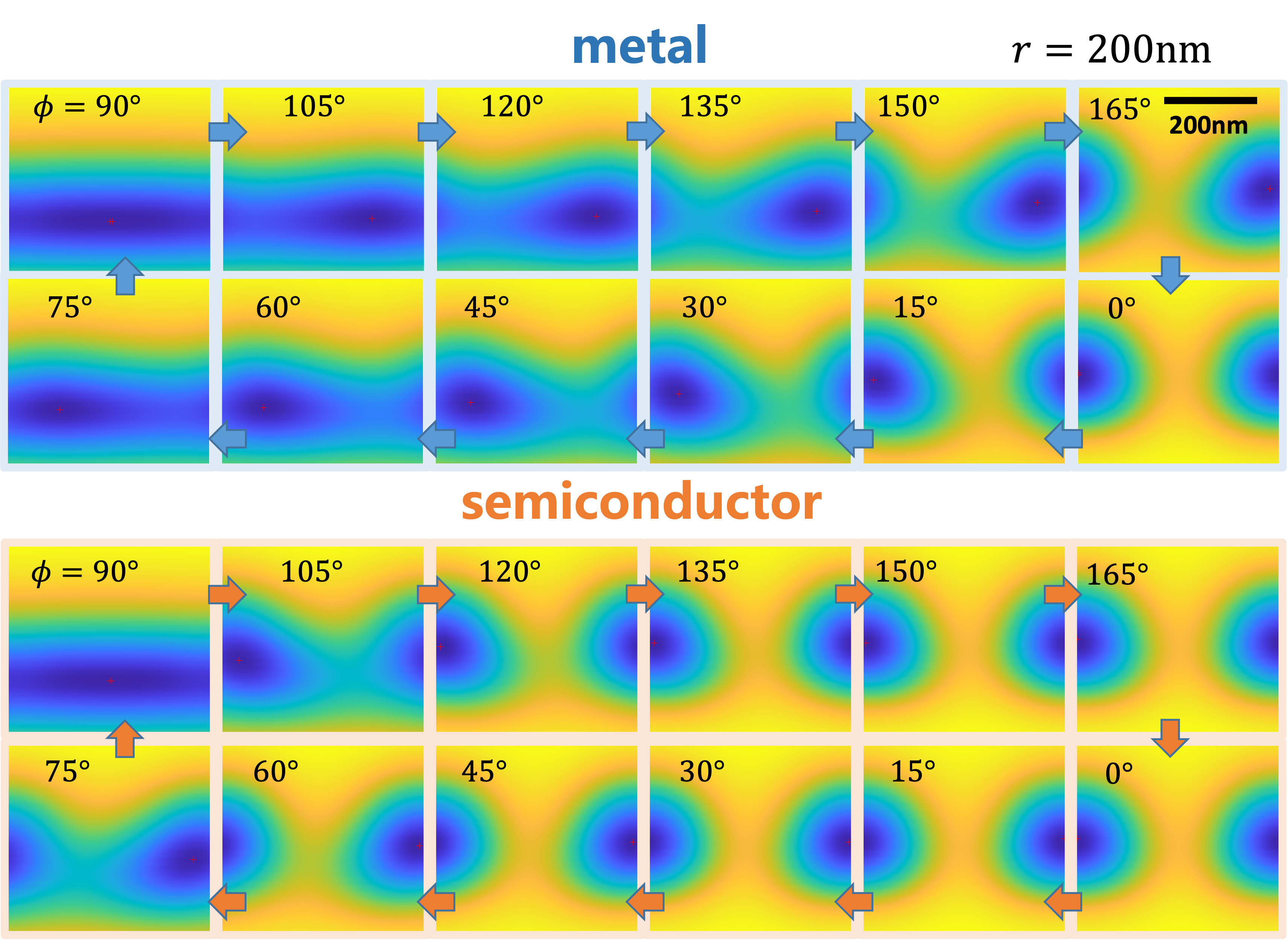}
\caption{The potential field of UI-belt for particle with $r=200\si{nm}$ in two \ce{VO2} phases and different polarization directions.}
\label{UI-belt-200}
\end{figure}

For a sphere with radius $r$, the force is an average of the forces on each part of the sphere, so the potential field needs to be averaged by integrating over the volume of the sphere. The distance between the lowest point of the sphere and the surface is taken as 20nm, and the potential field of $r=50\si{nm}$ and $r=200\si{nm}$ is shown in Fig.\ref{UI-belt-50} and \ref{UI-belt-200}.

\par
From these two figures we can see that the averaged potential field also has a moving bottom point. When $r=50\si{nm}$, the particle's transport is not very smooth due to its small size. But for $r=200\si{nm}$ the transport is more continuous. However some more larger particle may have a very mild potential field, which leads to a lower transport efficiency. The efficiency of optical conveying is highest only for particles within a certain range of sizes, which makes particle sorting in a specific size range possible.

\par
\begin{figure}
\centering
\includegraphics[scale=0.3]{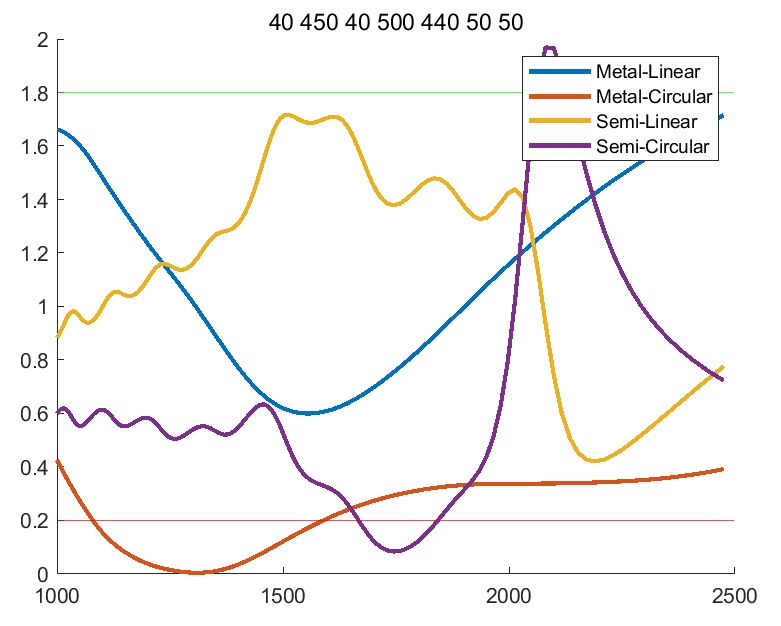}
\caption{The near-field intensity $I_{+P}$ of the prototype of UII-belt's unit structure at different phases and different incident polarization states. The real unit structure of UII-belt has its $L_1$ 100nm larger than the prototype, and two adjacent arms are merged into one.}
\label{UII-belt-single}
\end{figure}

Another type of the convey belt (named UII-belt) is to use polarization state with phase varying as the incident, i.e. the phase difference between $E_{y'}$ and $E_{x'}$ ($x'\&y'$ is $-45°$ rotated compared to the $x\&y$ coordinate) changes continuously from 0° to 360°, thus polarization shifts from x-polarized to LCP to y-polarized to RCP and so on. The phase difference is denoted as $\phi$. Similarly to UI-belt, the parameter of UII-belt is $h_1=h_3=40\si{nm},\ h_2=450\si{nm},\ L_1=600\si{nm},\ L_2=440\si{nm},\ w_1=50\si{nm},\ w_2=50\si{nm}$ (only $L_1$, and it is designed based on its prototype with a difference $L_1=500\si{nm}$. The near-field intensity of the prototype structure is shown in Fig.\ref{UII-belt-single}.

\begin{figure}
\centering
\includegraphics[scale=0.35]{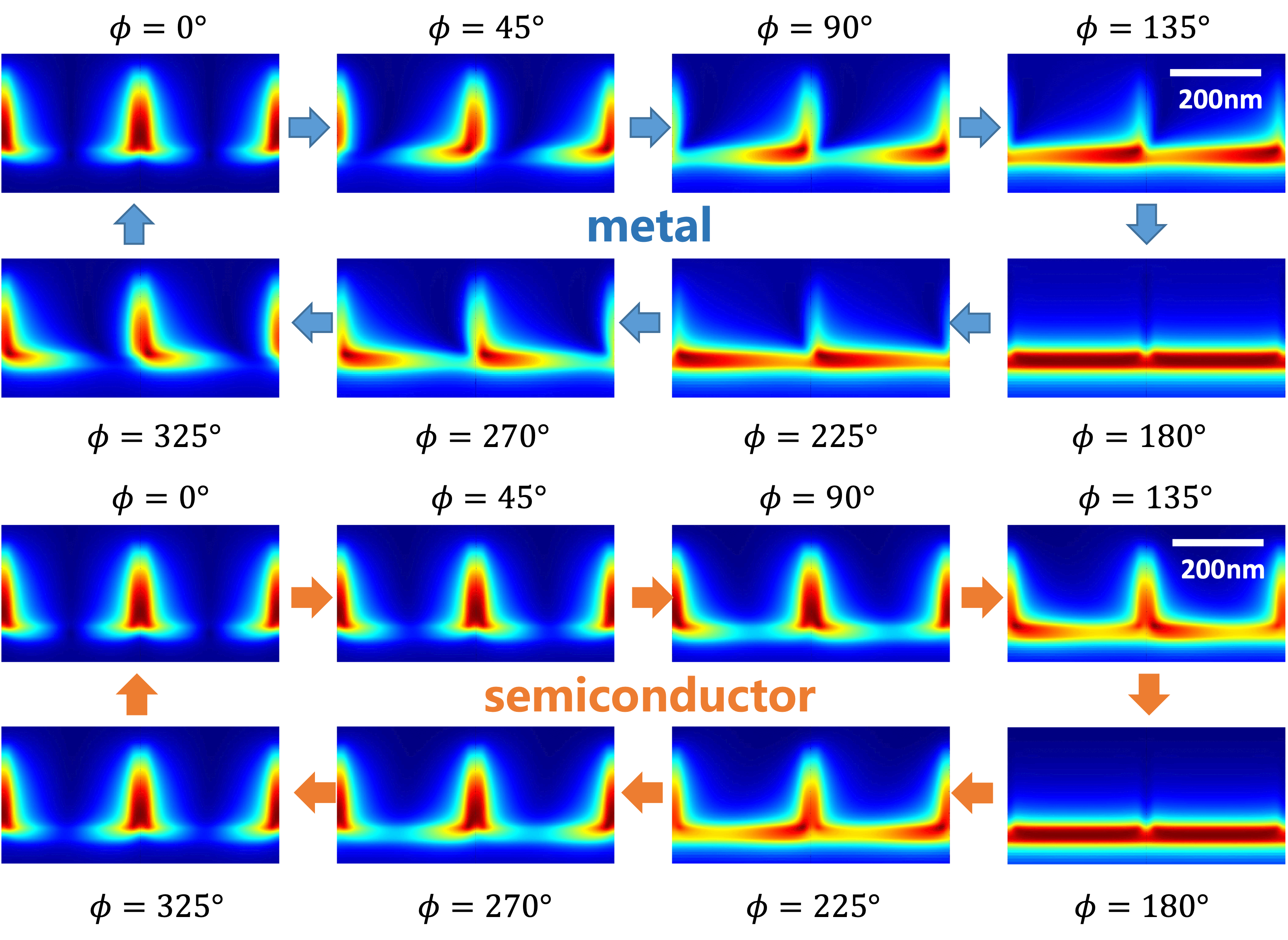}
\caption{The field distribution of UII-belt in two \ce{VO2} phases and different polarization directions, and $\lambda=2800\si{nm}$, $\phi$ is the phase difference between $E_{y'}$ and $E_{x'}$. The range of figure is two periods, and monitor plane is 20nm above the surface.}
\label{UII-belt-field}
\end{figure}

\begin{figure}
\centering
\includegraphics[scale=0.33]{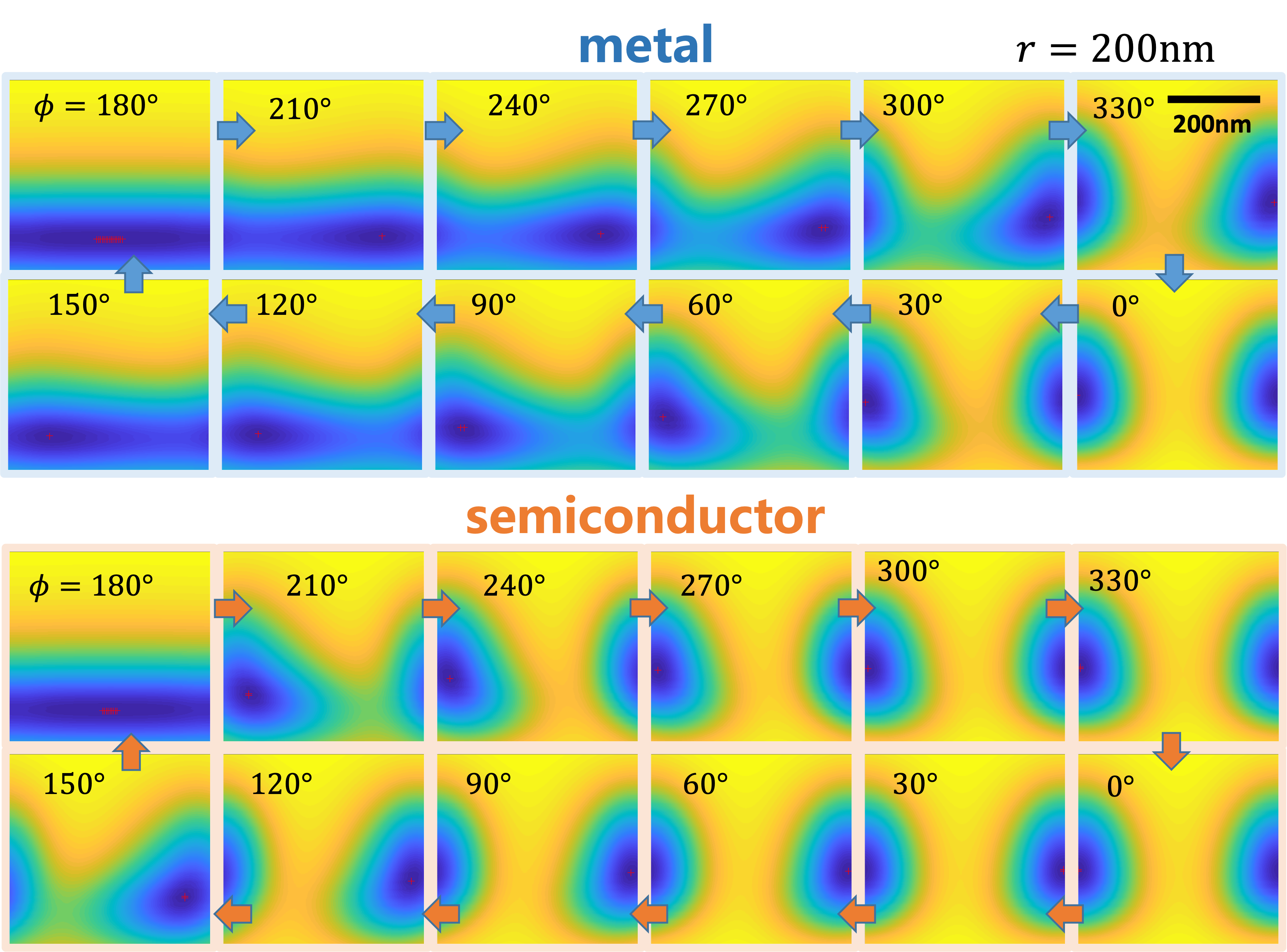}
\caption{The potential field of UII-belt for particle with $r=200\si{nm}$ in two \ce{VO2} phases and different polarization directions.}
\label{UII-belt-200}
\end{figure}

It has a peak for LCP in the semiconductive phase at about $\lambda=2100\si{nm}$ (2800nm when medium is water), while around this wavelength the curve in metallic phase has a low flat long tail. This can give the hot spot's transport, along with the averaged potential well for $r=200\si{nm}$, as shown in Fig.\ref{UII-belt-field} and \ref{UII-belt-200}.

\par
Similar to the UI-belt, the transport direction can also be inversed in different \ce{VO2} phase and the particle $r=200\si{nm}$ moves more smoothly as the polarization phase varying from 0° to 360°. 

\par
To constructing a conveyor belt that the peaks of intensity distribution flow from one unit to another with continuous variation of some parameters, we need a monotonic perodic variation. Hence adjusting the phase or angle of the polarization state is the best method, and with both of two ways we've designed the two U-belts that evolve from inversible linear or circular polarization splitter. It is worth noting that phase-varying elliptically polarized light has not been used as a conveyor belt incident source. The comb-like structure is also proposed for the first time as a conveyor belt,  in which these vertical arms play a key role of transfering the hot spots. Furthurmore, the nanofilm structure composed of \ce{Au-VO2-Au} also exhibits its excellent performance in regulating the transport direction. 

\section{Conclusion}

In conclusion, we have studied the SPP resonance modes of the U-shaped nanohole structure composed of \ce{Au-VO2-Au} film, obtained the basic SPP distribution characteristics in the near-field and far-field of the U-hole at different sizes, polarization states and material phases, and demonstrated the splitting effect of a multifunctional polarization splitter designed in this way. 

\par
By using the near-field distribution characteristics, we design a switchable double-ended output optical logic gate with advantages of nano-size and flexibility to switch different types of gates. 

\par
Through the far-field distribution characteristics, we also designed a bi-directional optical conveyor belt made by merging multiple U-holes, and we can well regulate the transmission speed and direction of nanoparticles through the change of polarization phase (or polarization direction) and material phase. This is a very promising optical tweezer structure.

\section{Acknowledge}
\begin{acknowledgments}
Thanks to the help of supervisor Jing Yang, and the calculation of averaged potential is helped by Yuxing Chen.
\end{acknowledgments}


\end{document}